# Polycrystalline Morphology and Anomalous Hall Effect in RF-Sputtered Co2MnGa Films

Carter E. Wade[1], Sunny Phan[2], Katherine Coffin[3], Gabe Paynter[2], Ty J. Cawein[1], Kurt G. Eyink[4], Andrei Kogan[2], Joseph P. Corbett[1,a)]

1Department of Physics, Miami University, Oxford, OH 45056, USA
2Department of Physics, University of Cincinnati, Cincinnati, OH 45221, USA
3Department of Chemistry, Kenyon College, Gambier, OH 43022, USA
4Materials and Manufacturing Directorate, Air Force Research Laboratory, Wright Patterson Air Force Base 45433
a) Electronic mail: corbetj5@miamioh.edu

The Heusler compound $Co_2MnGa$ is a topological semimetal with intriguing electronic and magnetic properties, making it a promising candidate for spintronic applications. This study systematically investigates the effects of substrate temperature and RF sputtering power on the structure, morphology, and anomalous Hall effect (AHE) in $Co_2MnGa$ thin films. Using X-ray diffraction line analysis, we identify variations in film orientation and crystallinity, revealing the emergence of high-index textures at specific growth conditions. Atomic force microscopy imaging provide insight into grain morphology and size distributions demonstrating a correlation between deposition parameters and film texture. Hall transport measurements confirm a strong dependence of AHE on growth conditions, exhibiting a non-monotonic relationship with RF power and temperature. Despite significant variations in microstructure, a striking linear relationship between AHE and the zero-field slope of the Hall resistivity is observed, suggesting an underlying universal mechanism. These findings provide a foundation for investigating the complex interplay of CMG thin film conditions and transport for next-generation magnetic and electronic devices.

## I.   INTRODUCTION

Topological semimetals [1] are a subclass of quantum materials that can arise under band inversion alongside large spin-orbit coupling and symmetry breaking in time reversal or centrosymmetric symmetries. Under these conditions, protected topological electronic states in the bulk and surface can develop [2], including Weyl points and nodal lines in the Brillouin zone in momentum space. These lead to a range of novel electronic responses as a function of temperature and magnetic field, observable not only at cryogenic but also at temperatures as high as room temperature, such as an unusually strong anomalous Hall effect [2-4] which makes these materials desirable candidates for practical spin-based sensors and logic elements. $Co_2MnGa$ (CMG) [5] is a metallic Heusler type ferromagnet [6] that has recently attracted significant attention due to theoretical predictions of half metallicity accompanied by Weyl nodal line regions in the Fermi surface. These features, resulting directly from the crystal structure of the material, have been shown in recent experiments to yield the appearance of unusual drum head topologically protected surface states [7] as well as strongly anisotropic surface transport in single crystal samples [8].

An important open question: How do the properties of ideal single crystals, observed or predicted, relate to the behavior of technologically relevant thin films grown by a range of sputtering methods that are usually of a polycrystalline form? This leads to a complex interplay of bulk and surface physics in the sample, which we begin to address in this work.

In this paper, we explore the relationship between the microscopic film morphology, local magnetic order, and macroscopic Anomalous Hall Effect (AHE) tuned by the substrate temperature and the radio frequency (RF) sputtering power during the film growth as a step towards improving control over the film properties beyond direct current (DC) sputtering techniques.

The fully ordered L2$_1$ CMG Heusler compound crystallizes in the cubic Cu$_2$MnAl-type structure with space group Fm$\underline{3}$m (No. 225) and lattice constant $a = 5.77$ Å. In this fully-ordered L2$_1$ phase, the Co atoms occupy the Wyckoff position 8c, whereas Mn and Ga occupy the positions 4b and 4a, respectively [9]. As shown in Figure 1, CMG crystallizes in the cubic structure with four interpenetrating face-centered cubic sublattices composed of A, B, C, and D sites. The A, B, C, and D sites correspond to (0, 0, 0), (1/4, 1/4, 1/4), (1/2, 1/2, 1/2), and (3/4, 3/4, 3/4) in the Wyckoff coordinates respectively. In this fully ordered CMG, Co atoms occupy the A and C sites, Mn the B site, and Ga the D sites. The B2 structure is similar to the L2$_1$ with the modification that the B and D (Mn, Ga) sites are chemically mixed. The A2 structure is a further mixing of the chemical species, where all sites can contain any element, resulting in a further reduction in crystal symmetry. CMG has been studied extensively as a bulk crystal and as a thin film with several magnetic and transport studies from direct current DC sputtered films. High-quality CMG films have been produced with DC sputtering, demonstrating an L2$_1$-ordering through triple-axis X-ray diffraction measurements. Sato et al. [10, 11] carefully studied the impact of film stoichiometry on the response of the AHE, finding a maximum in the AHE at the nominal stoichiometry. At the same time, Wang et al. [12] studied the impact of polytype and amorphous to partially crystalline CMG films as a function of temperature, where these studies focused on magnetometry and transport characterization uncovering strong AHE response up to 12.5% in L2$_1$-ordered CMG films with various degrees of AHE in B2, amorphous, and partially L2$_1$-amorphous films. Lacking in the literature are survey studies exploring the growth parameter space.

This study expands on prior work by systematically exploring the effects of the growth parameters on the CMG film morphology, growth orientation, and the anomalous Hall effect. Below, we present data from an array of films grown over a range of substrate temperatures and sputtering RF power input to a mosaic CMG target. We characterize the growth orientations, crystal quality, and texturization by utilizing diffraction line analysis alongside atomic force microscopy. We uncovered a complex parameter space for RF sputtered films, enabling the growth of unique high-index films with controllable crystallite size and shape. Room temperature transport measurements in magnetic fields up to 9 Tesla show a surprisingly strong variation of the anomalous Hall resistance with the growth conditions. We also observe magnetic stripe domains in the demagnetized domain structure across the parameter space using magnetic force microscopy (MFM) imaging.

## II. EXPERIMENTAL

All CMG film samples were grown via coil-assisted RF magnetron sputtering in a custom-built UHV sputter epitaxy system. The growth system consists of an UHV deposition chamber with a base pressure of $1\times10^{-10}$ Torr and an adjoining load lock with a $1\times10^{-8}$ Torr base pressure separated by a gate valve. Substrates are introduced into the system via a commercial, removable high-temperature heater stage provided by UHV Transfer, with substrates mounted directly onto the heater via molybdenum spring clips. Substrate temperatures were measured with a thermocouple sensor mounted on the removable substrate heater and correlated to the heater power via calibration. The removable heater transfers onto a 4-axis goniometer that provides an x y z and polar angle positioning relative to the axis of the sputter gun.

The 1.3" CMG sputter targets were produced by compressing elemental powders (~350-500 mess) in a 40-ton cold isostatic press. The constituent elemental powders were weighed to 0.001 g precision to achieve a CMG stoichiometry, and a 10 g target was pressed. Targets were pressed for 1 week at 40 tons, sintered at ~300 °C under vacuum (10-7 Torr) for a week, then powdered and repressed at 40 tons for a week. A pair of custom external Helmholtz coils were mounted coaxially to the sputter guns, allowing the sputtering plasma density to be increased (focused) or decreased (defocused), enabling additional control of the plasma density. The field generated by the Helmholtz coils was measured at the sample growth position via a Gauss meter.

## A.  *Sample growth*

This study encompasses a systematic variation of sputtering power and substrate temperature during film deposition. A total of 9 samples were grown on single-side polished 5×5×0.5 mm MgO(001) substrates spanning three temperatures: 330 °C, 420 °C 630 °C, and 3 RF powers at each temperature: 10 W, 25 W, and 50 W. Each sample was first brought to the intended deposition temperature, and 1 hour was allowed for thermal stabilization prior to deposition. Ultra-high purity Argon gas was introduced via a UHV leak valve approximately 20 minutes prior to deposition, during the temperature stabilization period, and regulated at 16 mTorr pressure during depositions. The plasma was ignited at the desired power while the sputter gun was shuttered, and a current of 30 A was applied to the Helmholtz coils, producing a focusing magnetic field of 21 mT. All samples were grown for 1 hour, resulting in a range of sample thicknesses as a function of power. All samples were allowed at least 1 hour to cool under UHV before being transferred through the load lock and removed from the heater for *ex-situ* characterization.

## B.    Characterization Techniques

Crystallographic, morphological, compositional, and magnetic characterization were performed using a combination of atomic force microscopy (AFM), X-ray diffraction (XRD), and energy dispersive X-ray analysis (EDX). A commercial high-resolution (14 pm) multi-modal AFM by AFMWorkshop was used for surface characterization and magnetic force imaging (MFM). Commercial Bruker MESP-V2 tips were used for MFM scanning. All samples were demagnetized with an alternating current demagnetizer prior to MFM studies. X-ray diffractometry was performed using a Bruker D8 Advance system with a Cu-K$\alpha$ X-ray source. Coupled $\theta$-2$\theta$ scans were used to determine the lattice constant, crystallinity, phase, orientation, texture, and coherence length present in the films. An Apreo scanning electron microscope with *in-situ* EDX was employed to determine the samples' stoichiometry with a 5 kV accelerating voltage. Deposition rates were determined using profilometry within the AFM, and nominal deposition thickness are calculated from these rates.

All transport measurements were performed at room temperature in a vacuum ($<10^{-5}$ mTorr) enclosure.  A cryocooled superconducting magnet (Cryomagnetics, Inc.) with a room-temperature bore was used to produce magnetic fields up to 9 Tesla. A DC transport circuit consisted of A Keithley 2400 sourcemeter, a DL instruments model 1211 current amplifier used to supply and record  a ~ 1 mA current through the sample (Figure 2 ) and an Ithaco 1201 voltage preamplifier to record a transverse voltage $V_{xy}$ on the sample. All samples measured were as-grown, 5×5×0.5 mm squares mounted on a custom-made copper-clad circuit board used to interface the sample to the system wiring.

Low-resistance electrical contacts to the samples were approximately 1 mm in diameter silver paint (Ted Pella Inc., Leitsilber 200) dots positioned on the opposing edges of the sample. The magnetic field direction is chosen to be positive for the field into the page, as shown in the figure, which gives a positive $R_{xy}$ for p-like transport. We have measured $R_{xy}$ as a function of B for all samples in a wide range of fields. A typical data (Figure 2, b). shows features characteristic of the anomalous Hall effect: the sharp field dependence at low fields and saturation at higher fields, see Figure 2(b). The following section presents the results of the sample characterization and transport measurements for the different growth regimes.

## II. RESULTS AND DISCUSSION

### A. Growth Orientations, Compositions, Texture Coefficients, and Degree of Crystallinity

Figure 3 shows XRD data for symmetrically coupled scans as a function of deposition temperature and sputter power. Systematic characterization of the film series is performed through a series of XRD line analyses: hkl indexing, texture coefficient ($C_{hkl}$), and degree of crystallinity (DoC). These films demonstrate polycrystalline growth but tend to have a significant preference for one or two orientations. We also characterize the composition of the films through EDX analysis. This reveals slight compositional variation depending on deposition temperature and sputter power. All films grown in the study are at the CMG stoichiometry within a few percent.

Identification of L2$_1$, B2, and A2 phases from symmetric coupled-scan XRD measurements can be challenging, especially in polycrystalline films where the three phases can coexist. Considering that we span a range of deposition temperature and power conditions, impacting order, this is a possible scenario for this work. The reduction of symmetry from a primitive B2 symmetry to a face-centered L2$_1$ symmetry causes extinction in numerous reflections, especially in the odd-ordered reflections for the h00 family (i.e., 100, 300, 500 reflections). However, to compound the complexity, accidental extinctions from compositional scattering factors cause numerous extinctions across all three phases. For example, the odd-ordered h00 reflections forbidden in L2$_1$ but permitted by symmetry in B2 are also extinct from compositional scattering factors. This creates a scenario where the body-centered A2 has the fewest number of reflections, B2 has a few than A2, and L2$_1$ has the most unique set of reflections, for example, the 111, 311, and 331 reflections–of these reflections, the 111 reflection is typically used in a triple-axis XRD measurement for L2$_1$ assignment of 100 oriented films.

For the deposition temperatures of 630 °C and 420 °C across all sputter powers (10 W, 25 W, 50 W), we observe a high-intensity even-order h00 family of peaks with low-intensity inclusions of high-order reflection. One exception is the deposition at 630 °C at 25 W, a pronounced L2$_1$ CMG 311 reflection is observed, indicating L2$_1$ crystallization. At really high angles (2θ ~120 degrees), a 533 peak is also pronounced. This specific growth has a more uniform strain compared to the other growths, as indicated by the h00 family with respect to MgO substrate peak alignment.

For the 330 °C deposition temperature, we do not observe the pronounced h00 family of peaks but instead observe the 110 and the 422 peaks alongside a peak that cannot be attributed to CMG. Instead, this peak is attributed to a Co peak of 101 orientation, for which we suspect the low deposition temperature enables the crystallization of Co inclusions. Using the observed reflections from our XRD measurements, the CMG lattice constants are determined as a function of the deposition temperature and power and plotted in Figure 4. We observe a very minimal lattice variation for our combination of deposition temperature and power where lattice constants agree with the 5.77 Å bulk value to within ±0.01 Å, indicating minimal macrostrain present in the films. A deviation from this overall trend is the 50 W, 630 °C condition, where a larger lattice constant of 5.86 Å is observed.

Since we have polycrystalline films with pronounced intensities of certain families of peaks, we utilize texture coefficient analysis to ascertain which orientation is the predominant film orientation, enabling a quantitative measure of preferential film orientations normalized to random distribution of orientations (i.e., powder diffraction intensities). In thin films, the measured XRD intensity is a combination of the number of scattering crystallites and the scattering amplitudes of the material for the given hkl reflection. Since the texture coefficient is normalized to the random distribution intensities, this accounts for differences in intensity brought on by scattering amplitudes while simultaneously enabling a comparison of intensity between peaks of a measured sample. The texture coefficient is calculated via the formula:

$$C_{(h_i k_i l_i)} = \frac{I_{(h_i k_i l_i)}}{I_{0,(h_i k_i l_i)}} \left( \frac{1}{N} \sum_{i=1}^{N} \frac{I_{(h_i k_i l_i)}}{I_{0,(h_i k_i l_i)}} \right)^{-1}, \quad [1]$$

where $I_{hkl}$ is the integrated intensity of a given diffraction peak corresponding to a Miller index hkl, $I_{0,hkl}$ is the relative integrated intensity of the same hkl peak of randomly orientation diffraction pattern, and $N$ is the number of sample peaks considered in the analysis. The texture coefficients can take values of $0 \leq C_{hkl} \leq N$. A $C_{hkl}=1$ indicates a crystallite presence equivalent to that found in a randomly oriented sample.

For $C_{hkl} > 1$, this indicates a preferential orientation in the film, whereas a $C_{hkl}=N$ indicates a single orientation. For $0 < C_{hkl} < 1$ indicates sparse inclusions in the film growth. We perform $C_{hkl}$ calculation across the series of samples spanning deposition temperature and sputter power; see Figure 5. We group $C_{hkl}$ values into groups of reflection families (e.g. h00), and by phase type CMG or Co. For the lowest deposition temperatures, a two-phase system of CMG and Co is formed; addressing this within the $C_{hkl}$ analysis is straightforward since $C_{hkl}$ is normalized to a random distribution of crystallites without preference for CMG or Co. We utilize PowderCell to simulate powder diffraction intensities of both CMG and Co phases and normalize the spectrum to the highest intensity to serve as our reference intensities.

Figure 5 shows a bar chart of $C_{hkl}$ across the span of deposition temperature and powers investigated. Solid colors indicated CMG phases, while dotted colors indicated Co phases. In the 330 °C deposition temperature, we see Co dominating the film texture at 10 W, whereas at 25 W, CMG 220 oriented films become more prevalent until 50 W, where CMG 111 oriented films dominate the growth. Then, a marked transition occurs with temperature, where for 420 °C growths across all powers, the 001 orientation becomes the dominant texture with 002 (light green bar), 004 (brown bar), and 006 (dark green) bar. Unexpectedly, at 630 °C at 25 W, higher order phases become prevalent, particularly the 533 orientation (purple bar) and 311 (red bar). Where at 630 °C and 25 W, the 533 and 311 orientations dominate.

While the texture coefficient gives a more quantitative measure of orientation preference, it does not consider the degree to which the films contain amorphous material. Amorphous material does not contribute to narrow diffraction peaks but rather a broad background hump in the data. By systematically accounting for X-ray signals from the diffractometer optics, sample holder, and substrate, a degree of crystallinity can be determined by analyzing the integrated intensity of the entire diffractogram. See supplemental Figure S1 for a comparison of the sample holder and substrate alone. To address this within our series, we compute a degree of crystallinity (DoC) for each sample from the ratio of the integrated intensity of all crystalline sample peaks to the total scan area of the instrumental and substrate background-subtracted total scan area. Which is given in the formula:

$$DoC \ (\%) = \frac{\sum_{i=1}^{N} I_{(h_i k_i l_i)}}{\int_{2\theta} dI_{tot}(2\theta) - \int_{2\theta} dI_{0BG}(2\theta) - \sum I_{substrate}} \times 100\%, \ [2]$$

where $I_{hkl}$ is the integrated intensity of a given diffraction peak corresponding to a Miller index of the CMG sample, $\int dI_{tot}$ is the total integrated area of a sample scan without correction, $\int dI_{BG}$ is the total integrated area of a background scan of the sample holder, and $\sum I_{substrate}$ is the sum of the integrated intensities of all diffraction peaks in the sample scan due to the MgO (001) substrate (including the 200 and 400 reflections from Cu $K_{\alpha 1}$, Cu $K_{\alpha 2}$, Cu $K_{\beta 1}$, Cu $K_{\beta 2}$, W $L_{\alpha 1}$, and W $L_{\alpha 2}$).

Figure 6 shows the DoC as a function of sputter power for the three temperature series. Interestingly, we observe a non-monotonic trend where the highest temperature and highest power deposition conditions do not produce the most crystalline thin films. Instead, a combination of balancing the impinging energy of the sputter material and deposition temperature achieves a degree of crystallinity of up to 80%. For the lowest deposition temperature, a very low degree of crystallinity was observed, down to ~20%, even across the varying power.

Interestingly, if one combines the $C_{hkl}$ and DoC analysis, a more complete picture of the film is achieved. While the lowest deposition temperature has Co inclusions alongside CMG orientations, it is markedly not very crystalline. To make a more robust measure of our films, we combined the $C_{hkl}$ and DoC methods to create a crystallinity texture coefficient–which is simply the product of the two quantities scaled by *N*. With this new method, films that are highly crystalline and contain only a single growth orientation will result in a crystallinity texture coefficient of 1, whereas a random distribution of polycrystalline grains (i.e., powder diffraction), the crystallinity texture coefficient results in a value of 1/N for each reflection. In essence, the crystallinity texture coefficient rescales $C_{hkl}$ to determine the proportion of the film that is crystalline.

Figure 7 is a bar graph of the crystallinity texture coefficient across deposition temperature and power. To include an indication of the amorphous material content, an additional bar of $1\text{-DoC}\times C_{hkl}/N$ is included (dark blue). Examining this new crystallinity texture coefficient, we find the data separates more cleanly into groups scaled by their DoC, see Figure 7. We concluded from this combination of analyses that there is an optimal temperature and power window to achieve the highest crystallinity while the specific temperature and power are selected for the hkl orientation of the film. Where it seems mid-power and temperature are optimal for achieving 001 film orientation with the lowest amorphous content, while high power and high temperature, high-index films are produced with low amorphous content.

## B.  *Debye-Scherrer Analysis*

Building from XRD analysis based on peak position and intensity, more information regarding the film quality can be ascertained from a careful analysis of peak widths. For a given diffraction peak, the peak width, or more pointedly, the integral breadth of the peak, contains information about the coherent X-ray scatterer size (alternatively referred to as crystallite size) and the residual strain of the particle. We determine the crystallite size and residual lattice strain of the CMG samples using the Debye-Scherrer Analysis. Two predominant terms contribute to the peak broadening of the scattering size and the microstrain; each of these components contributes to the different shapes of the profile. The broadening attributable to size effects contributes to a Lorentzian profile, whereas microstrain contributes to a Gaussian profile.

As such, a Voigt profile (a convolution of Lorentzian and Gaussian functions) was fit to each peak in the data using Fityk, from which we recorded the center, height, area, and Gaussian and Lorentzian full-width half-maxima (FWHM), which were converted into integral breadth via the formulae $\beta_G = (FWHM_G/2)\sqrt{\pi/\ln(2)}, \beta_L = (FWHM_L/2)\pi$. The integral breadth consists of three contributions: the instrumental broadening $\beta_{instrument}$, the crystallite size broadening $\beta_{size}$, and the lattice strain broadening $\beta_{strain}$. The instrumental broadening was determined using a corundum standard, which is taken to have negligible scatterer and strain broadening compared to the instrumental broadening and fit to a quadratic function to extract the angular dependent broadening.

To separate the size broadening from the strain broadening, the scatterer/crystallite size broadening $\beta_{size}$ and the lattice strain broadening $\beta_{strain}$ to be independent of each other and sum to give the sample broadening at a given diffraction angle. We perform single peak analysis by fitting a Voigt function to separate the components and of the broadening, and then solve for the size or strain utilizing the Scherrer and Wilson formulae: $\beta_{size} = \frac{K\lambda}{D\cos\cos(\theta)}$, $\beta_{strain} = 4\varepsilon\tan\tan(\theta)$, where $K$ is the shape factor (K=0.89 for our cubic materials), $\lambda$ is the wavelength of X-rays in the XRD ($\lambda$=0.154nm for Cu-Kα radiation), $D$ is the crystallite size, and $\varepsilon$ is the microstrain of the crystallite.

Figure 8 shows a plot of crystallite size and microstrain as a function of deposition power for the three deposition temperatures. Not surprisingly, for the coldest growths, we observe the smallest crystallite size of ~40 nm with an increase to ~100 nm at 50 W deposition power. For the mid-temperature series, a linear increase from ~40 nm crystallite size to 150 nm size at 50 W. The largest crystallite sizes were observed from the hottest deposition temperature that grew linearly to ~250 nm in size. Whereas the microstrain as a function of deposition power across the temperature series has a slight downward trend of 0.2% to 0.15%, see Figure 8.

## C. *Morphology and Grain Analysis*

Correlating morphology to growth conditions offers spatially resolved insights into the films, serves as a complement to XRD line analysis techniques, and further informs transport interpretations on the state of the film. Atomic force microscopy was used to measure the topographic surface features of all films grown with Gwyddion, which was used to process the data and perform statistical analysis. Figure 9 shows an array of AFM images with corresponding line cuts spanning the range of growth temperature and sputter power for the same sequence in the XRD analysis. Distinct trends in morphology have emerged as a function of deposition temperature and power, where the coldest growths of 330 °C displayed a very granular morphology with circular and rounded grains. These grains grew in characteristic size as a function of sputtering power from 10 W with a grain width of ~120 nm diameter to 50 W with a grain size of ~250 nm diameter. Based on our XRD analysis, we know these films are mostly amorphous, with the inclusion of crystalline materials within them. Compared to the Debye analysis of the inclusions, the grain size and correlation length of the crystalline inclusions are comparable (see Figure 8), which suggests that the grains that are crystalline are roughly the size of the average rounded grain observed in AFM. For the 420 °C depositions, a similar circular grainy morphology was observed, but the grains were significantly large in diameter ~400 nm and at higher deposition powers, with some grains displaying a relatively flatter surface. At the highest deposition temperature, a marked change in morphology occurs, where large flat terraces are observed. Comparatively, smaller terraces occur at low deposition powers, whereas at higher deposition powers, large flat terraces appear. Decorations of small, rounded islands are scattered atop these terraces and likely leftover material that has yet to be incorporated into the terrace morphology. Compared to our DoC calculations, these films deposited at 630 °C along with the 420 °C series, have the highest crystallinity content; however, it is clear that to achieve a flat terraced morphology, high temperatures to increase diffusion lengths are needed to smooth the terrace surface. However, a trade-off in morphology occurs across these growths; as the temperature and power increase, so does the grain size. However, the surface height variation also grows significantly, up to 200-hundred-nm variation.

We perform a statistical grain analysis on the array of AFM images above to ascertain quantitative measures of the observed morphological trends. We utilized native procedures within Gwyddion to mask-off grains by either a watershed or segmentation algorithm and compute statistical distribution from the mask. Similar to the morphological trends observed above in line cuts, the low temperature 490 °C growths show a clear delineation of grain height as a function of power, where the grain height grows with power. Along the same vein, the height distribution of these low-temperature grains also grows with power. Figure 10 (a,b) shows the collective trends determined from the statistical grain analysis for both grain (or terrace) height versus power and grain diameter (or side length) versus power. The corresponding statistical distributions from which the grain information was obtained are shown in Figure 10. As a general trend in the films, as temperature and power increase, the height variation of the surface and the grain (terrace) size increases.

## D.  Magnetic Domain Characterization

Figure 11 shows the demagnetized MFM data spanning the same temperature and power series as above, with a line cut across a domain in each image. Correspondingly, fast Fourier transforms (FFT) of the domain structures were computed to extract an average domain size from a Fourier analysis of the FFT pattern diameter (Figure 12).  The diameter of the FFT patterns is marked with a green circular ring within the FFT image and as green vertical lines within the corresponding line cuts. We see, broadly speaking, a striped phase in all films; the specific domain pattern varies as a function of temperature and power, but the domain sizes remain ~1 µm in width, see Figure 13. Magnetic domain characterization is performed utilizing line cuts of stripe domains to estimate domain size. We observe variation in the domain size as a function of deposition temperature and power, whereas a general trend is that the high-power films tend towards larger demagnetized domain sizes. For the mid-temperature deposition series, a meandering strip pattern is observed where the domain size increases as a function of RF power.

## E.   Hall Resistivity measurements

The Hall voltage was measured in each of the 9 samples as function of the magnetic field applied orthogonally to the sample plane. A DC current supplied from a Keithley 2400 SourceMeter entering the sample was measured, and compared to the current returned from the sample to the circuit ground via a current amplifier (Ithaco 1211), which provided a reading for the current exiting the sample. In all measurements, care was taken to ensure that the currents entering and leaving the sample are identical, to rule out any unwanted current paths from the sample to the apparatus ground. The transverse voltage $V_{xy}$ was measured via a low-noise preamplifier (DL Instruments 1201) and was digitized and recorded in a computerized data acquisition setup. The magnet current was simultaneously recorded via GPIB polling of the magnet current source. The field values presented here were obtained from the manufacturer's supplied current-field model (Cryomagnetics, Inc). We converted the Hall resistance $R_{xy} = \frac{V_{xy}}{I}$ (Figure 2, b) to resistivity $\rho_{xy} = R_{xy} \cdot d$, where $d$ is the film thickness, d= 0.16 mm, 1.10 mm and 2.00 mm for films grown at 10 W, 25 W and 50 W RF power, respectively.

Figure 12, a) shows resistivity measurements in all 9 samples. All films show a rapid increase of the resistivity at low fields followed by a saturation at  a nearly constant value $\rho_{AHE}$, characteristic of the Anomalous Hall effect with Hall polarity corresponding to a p-type transport. The data show that $\rho_{AHE}$,  the amplitude of the effect, varies with the growth conditions.  In the films  grown at the lowest two temperatures, 330 °C and 420 °C,  The values $\rho_{AHE}$ are consistently larger than in the films grown at 630 °C.  We find that $\rho_{AHE}$ varies with RF growth power non-monotonically. In films grown at 330 °C and 420 °C, the Hall resistivities $\rho_{AHE}$  obtained with  10 W and 50 W are comparable,  about 8 mW·cm$^2$

, but increases by almost a factor of 2 at mid power, 25 W. At 630 ℃, the peak in $\rho_{AHE}$ is not observed. Instead, we find comparable $\rho_{AHE} \sim 4$ mW·cm$^2$ at 10 W and 25 W RF power, and a surprising dramatic reduction of $\rho_{AHE}$, by more than an order of magnitude, in the film grown at 50 W.

Next, we discuss the variation of the shape of the $\rho_{xy}(B)$ curves with the growth conditions. The shape is characterized by three interrelated quantities: The maximum Hall resistivity $\rho_{AHE}$, the zero-field limit slope $d\rho_{xy}/dB$ and the saturation field $B_0 = \rho_{AHE}/(d\rho_{xy}/dB)$, which can be understood as an estimate of the magnetic field at which the initial linear increase in $\rho_{xy}(B)$ reaches $\rho_{AHE}$ (Figure 12b, inset).

The plot of $\rho_{AHE}$ versus the slope $d\rho_{xy}/dB$ at zero-B limit, Figure 12b, indicates that the data from all 9 samples are well described by the same linear function. This suggests that while $\rho_{AHE}$ can be tuned by almost 2 orders of magnitude by varying the growth conditions, the saturating field is the same in all films within the measurement errors.

With the exception of the 50W, 630C growth, Hall resistivities found in our measurements are quantitatively close to those obtained by other groups in much thinner films. Sumida et. al. [13] found $\rho_{AHE} \sim 7$-$20$ mW·cm$^2$ in 50 nm films grown at room temperature, with the higher resistivity value near 2:1:1 stoichiometry. Marcou et. al. [14] reported a thickness study, near the 2:1:1 composition and found $\rho_{AHE} \sim 11$-$14$ mW·cm$^2$ in films spanning a 20-80 nm range, which is also comparable to the $\rho_{AHE} \sim 6$-$16$ mW·cm$^2$ observed by Wang et. al. [12] in 30 nm films. We note that Wang's results indicated an increase in $\rho_{AHE}$ with growth temperature, whereas in our data, $\rho_{AHE}$ obtained from 630 ℃ growths is lower than from 330 ℃ and 420 ℃ growths. On the other hand, values as low as 2 mW·cm$^2$ were reported by Tong et. al. [15] in extremely thin (~6 nm) films grown via molecular-beam epitaxy. Zhu et. al. [16] investigated changes in $\rho_{AHE}$ with the structural film order and found values ranging from 0.5 mW·cm$^2$ to 11 mW·cm$^2$ with the lowest $\rho_{AHE}$ values observed in B2-ordered films grown at room temperature, and an increase in $\rho_{AHE}$ observed in films with L21 order.

Overall, the morphology, grain size, and orientation of the film are found to be just as important as the previously investigated features of crystallinity and allotrope with a variations by a factor of 10 in AHE found in our work. While trends emerge when connecting growth conditions to a measure property (DoC, C$_{hkl}$, D, etc.) optimizing the $\rho_{AHE}$ in thin films is more complex than maximizing thin film conditions. For example, 10 W 330 °C films which are ~90% amorphous CMG with terrace sizes ~40 nm, and 25 W 420 °C which are ~90% crystalline (presumably B2) with terrace sizes ~150 nm have comparable $\rho_{AHE}$. This lays the ground work for a much deeper dive into disentangling the structure-property relationship in CMG in thin film form.

## III.    SUMMARY AND CONCLUSIONS

In conclusion, we conducted a systematic study of growth conditions in magnetron sputtering of CMG as a function of deposition temperature and sputtering power and investigated impacts on film texture, morphology, and Hall transport. Thin films were characterized through XRD line analysis methods uncovering trends in texturization and crystal quality, where unusual high index orientations can be achieved. AFM imaging corroborated trends observed in XRD line analysis and revealed morphological trends not captured by XRD methods. Hall transport uncovered interesting trends in the AHE. Direct measurements show non-monotonic trend versus growth conditions, whereas a careful analysis of the AHE data revealed a strikingly linear trend of $\rho_{AHE}$ vs. d$\rho_{xy}$/dB spanning an order of magnitude in $\rho_{AHE}$ across our films. Whereas saturation field $B_0$ and domain size remains robust despite film variations. We conclude the control over film morphology, growth orientation, and grain size from the growth parameters, while the performance of the AHE depends significantly on the factors of morphology grainsize, crystallinity, allotrope and film orientation with as much as an order of magnitude variation in AHE. This work prompts further study into disentangling the relationship between AHE and the film properties particularly in high-index films and developing the linear trend of AHE vs . $d\rho_{xy}/dB$.

# ACKNOWLEDGMENTS

This material is based upon work supported by the National Science Foundation under Grant No. 2328747. This research was supported in part by the Air Force Research Laboratory Materials and Manufacturing Directorate, through the Air Force Office of Scientific Research Summer Faculty Fellowship Program®, Contract Numbers FA8750-15-3-6003, FA9550-15-0001 and FA9550-20-F-0005

# AUTHOR DECLARATIONS

**Conflicts of Interest**

The authors have no conflicts to disclose.

**Author Contributions**

**Carter Wade:** Formal Analysis(Equal) ; Methodology (Equal); Software (Lead); Validation(Lead); Visualization(Equal); Writing/Original Draft Preparation(Lead);

**Sunny Phan:** Formal Analysis(Equal); Methodology(Equal); Validation(Equal); Visualization(Equal); Writing/Review & Editing (Equal)

**Katherine Coffin:** Formal Analysis(Equal); Methodology(Equal); Validation(Equal) Writing/Review & Editing (Equal)

**Gabe Paynter:** Formal Analysis(Equal); Methodology(Equal); Validation(Equal) Writing/Review & Editing (Equal)

**Ty J. Cawein:** Methodology(Equal); Writing/Review & Editing (Equal)

**Kurt Eyink:** Conceptualization(Equal); Funding Acquisition (Equal); Writing/Review & Editing (Equal)

**Andrei Kogan:** Conceptualization(Equal); Formal Analysis (Equal), Methodology(Equal); Project Administration (Equal); Resources (Supporting); Writing/Review & Editing (Equal)

**Joseph Corbett:** Conceptualization (Lead); Formal Analysis (Equal); Funding Acquisition(Lead), Methodology(Lead); Project Administration (Lead); Resources (Lead); Software (Supporting); Supervision (Lead); Validation (Supporting); Visualization (Lead); Writing/Original Draft Preparation(Supporting); Writing/Review & Editing (Lead)

## DATA AVAILABILITY

Data available on request from the authors.

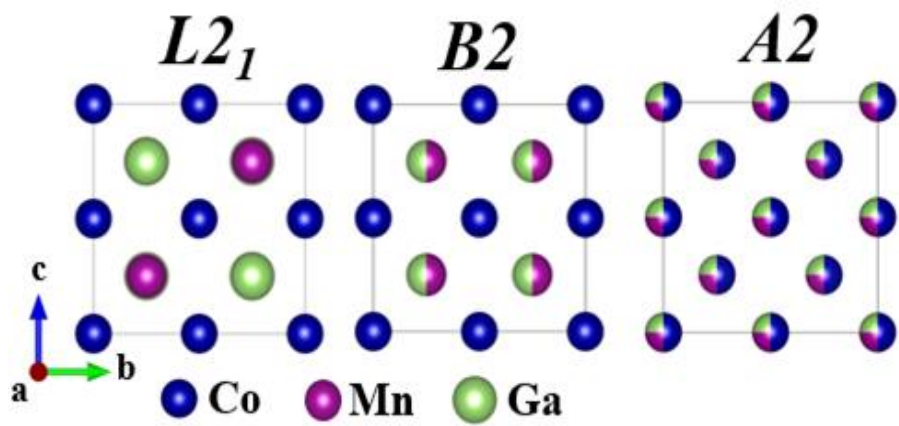

Figure 1 Crystal models of the three allotropes of CMG. Partially filled spheres represent occupancy. L2$_1$ ordering has the highest symmetry with a checkboard ordering of Ga and Mn (left). B2 ordering has a partial disorder with Mn and Ga sites becoming mixed with equal probability of occupancy as specified by the half filed green-purple spheres (middle). A2 ordering only preserves the geometric array, while having complete chemical disordering of the sites as specified by the partially filed sphere with the filing representing likely hood of occupancy (right).

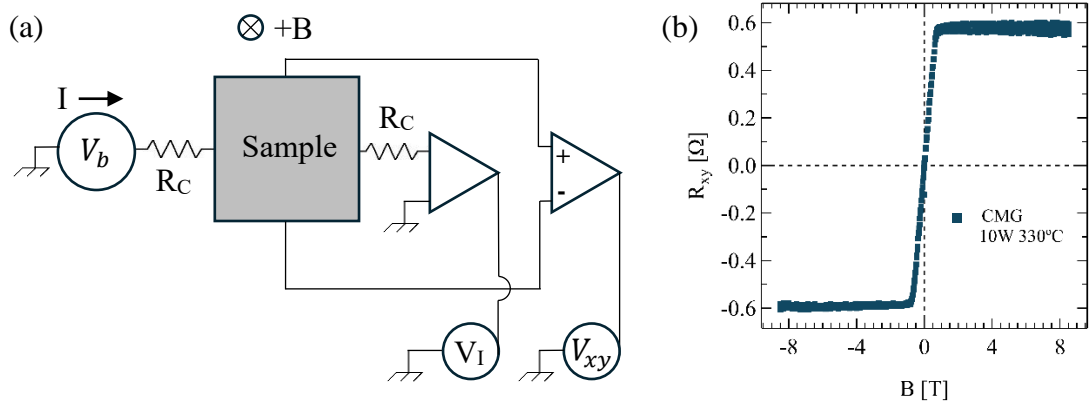

Figure 2 (a) DC transport circuit for measuring Hall signal. Sample (grey square) is connected to source $V_b$ and ground via two Ohmic contacts $R_C$ showing the position of the electric contacts for electric current I in the sample and transverse voltage Voltage $V_{xy}$ measurements. (b) Representative Magnetic field B [T] dependence of transverse resistance $R_{xy} = V_{xy} / I$ of CMG grown at 10W, 330°C.

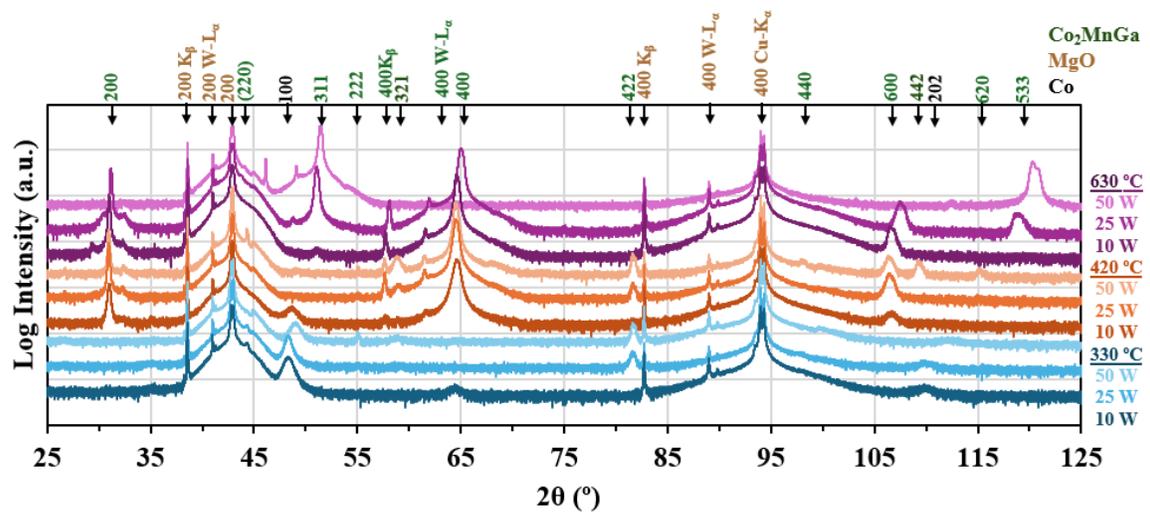

Figure 3 X-ray diffraction waterfall plot spanning the power and temperature series. Peak indexing of diffraction lines are indicated by the arrows with hkl values. Plots are color-coded with temperature and power, while hkl values are color-coded to source of reflection. Additional reflections from X-ray besides $K_\alpha$ are indicated.

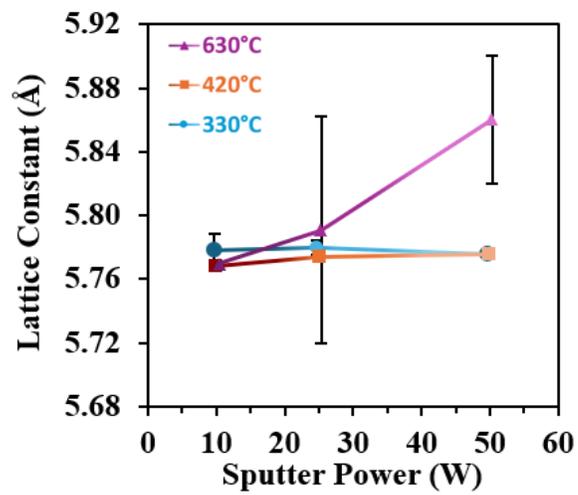

Figure 4 plot of lattice constant versus sputter power for the three temperature series as computed from X-ray diffraction peak positions.

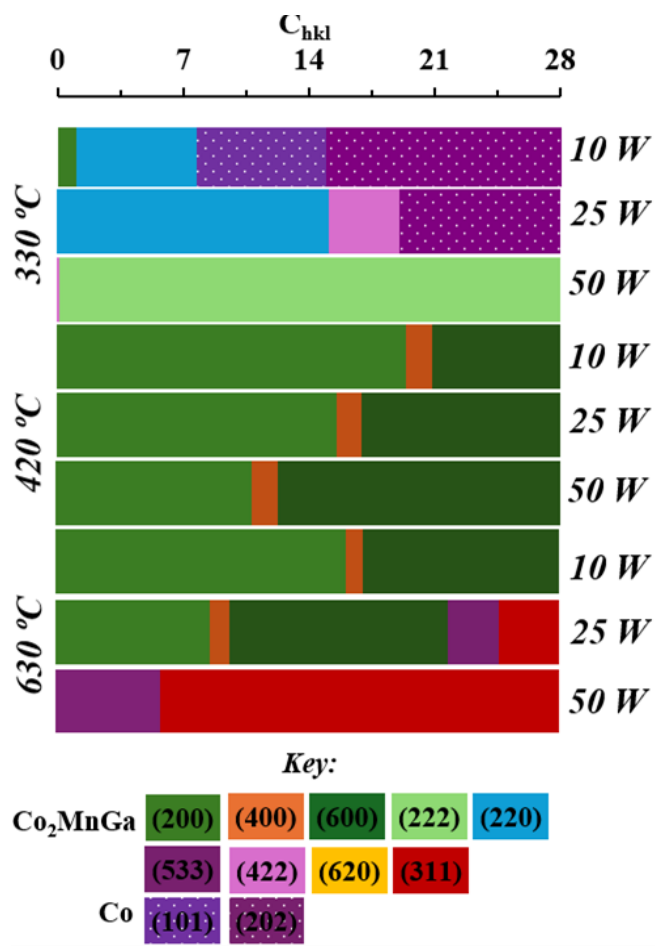

Figure 5 Texture coefficient analysis extract from XRD intensities spanning the temperature and power series.

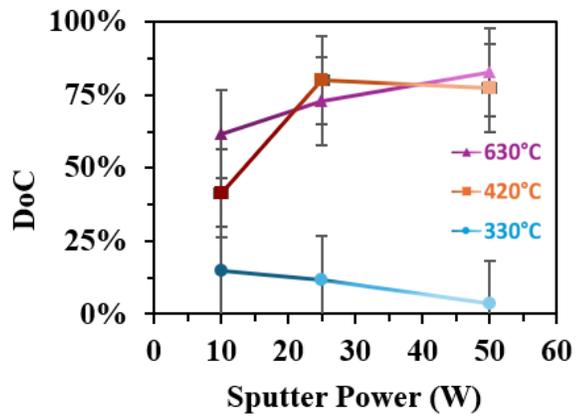

Figure 6 Degree of crystallinity as a function of sputtering power for the three temperature series.

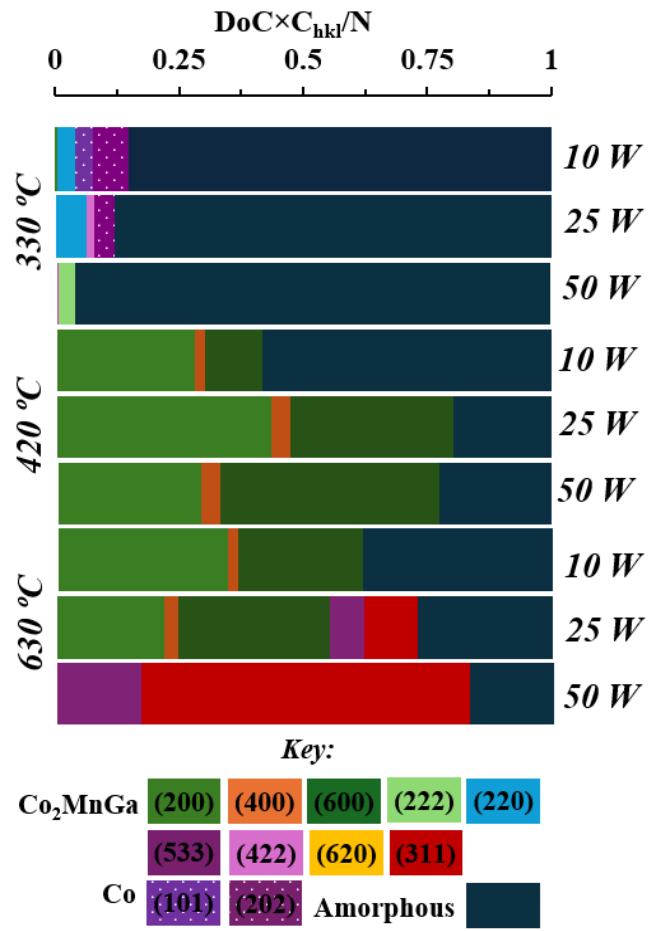

Figure 7 Crystalline texture coefficient analysis of the temperature and power series.

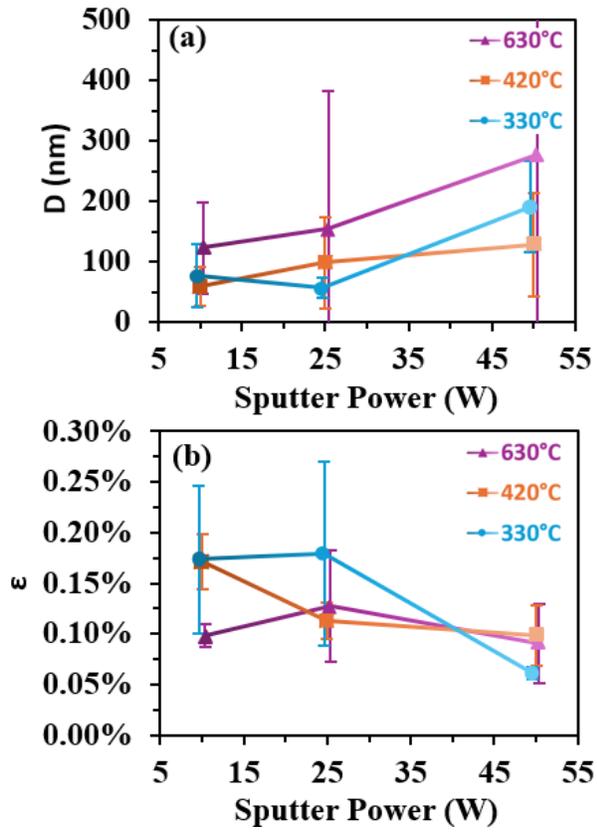

Figure 8 Debeye method to estimate X-ray coherence lengths and residual microstrain spanning the temperature and power series.

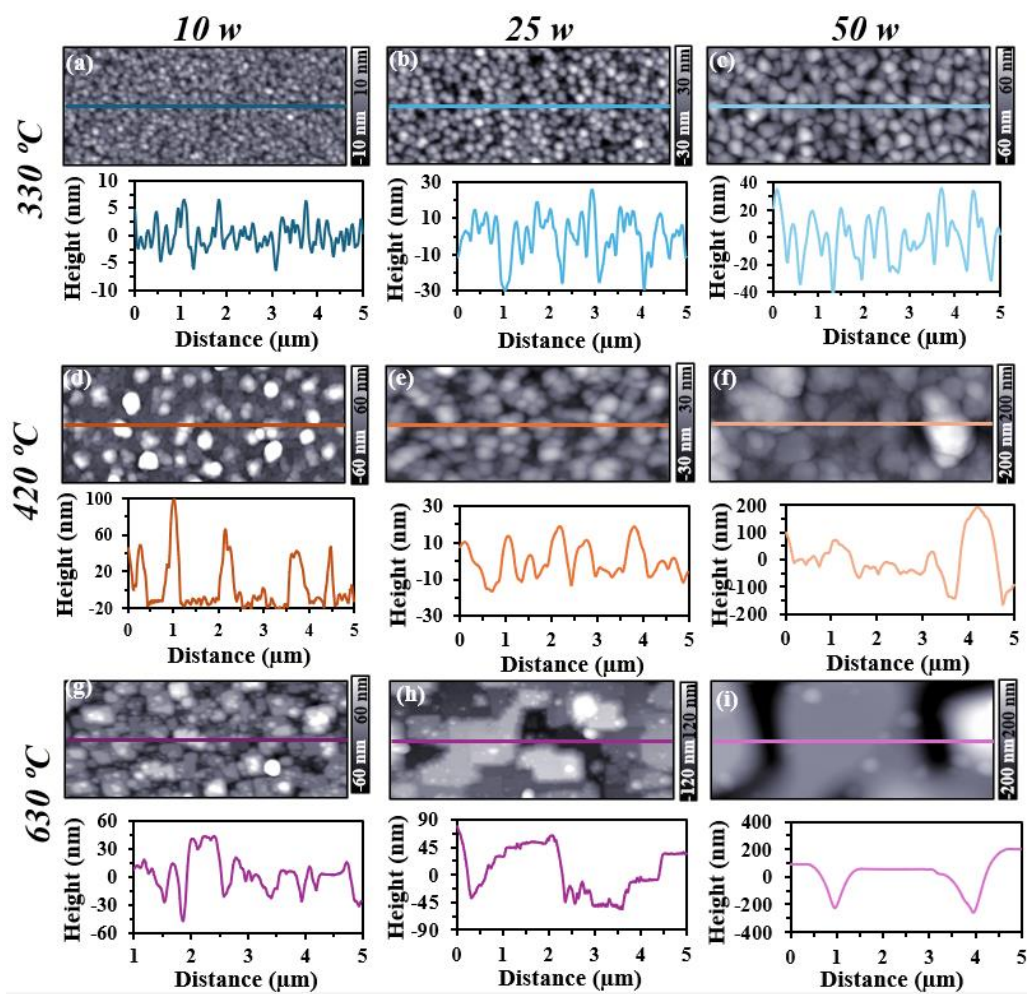

Figure 9 AFM topographic imaging for samples prepared as a function of power and temperature. Line cuts are color coded to each subpanel beneath the image. Trends in morphological control and growth are visualized with imaging.

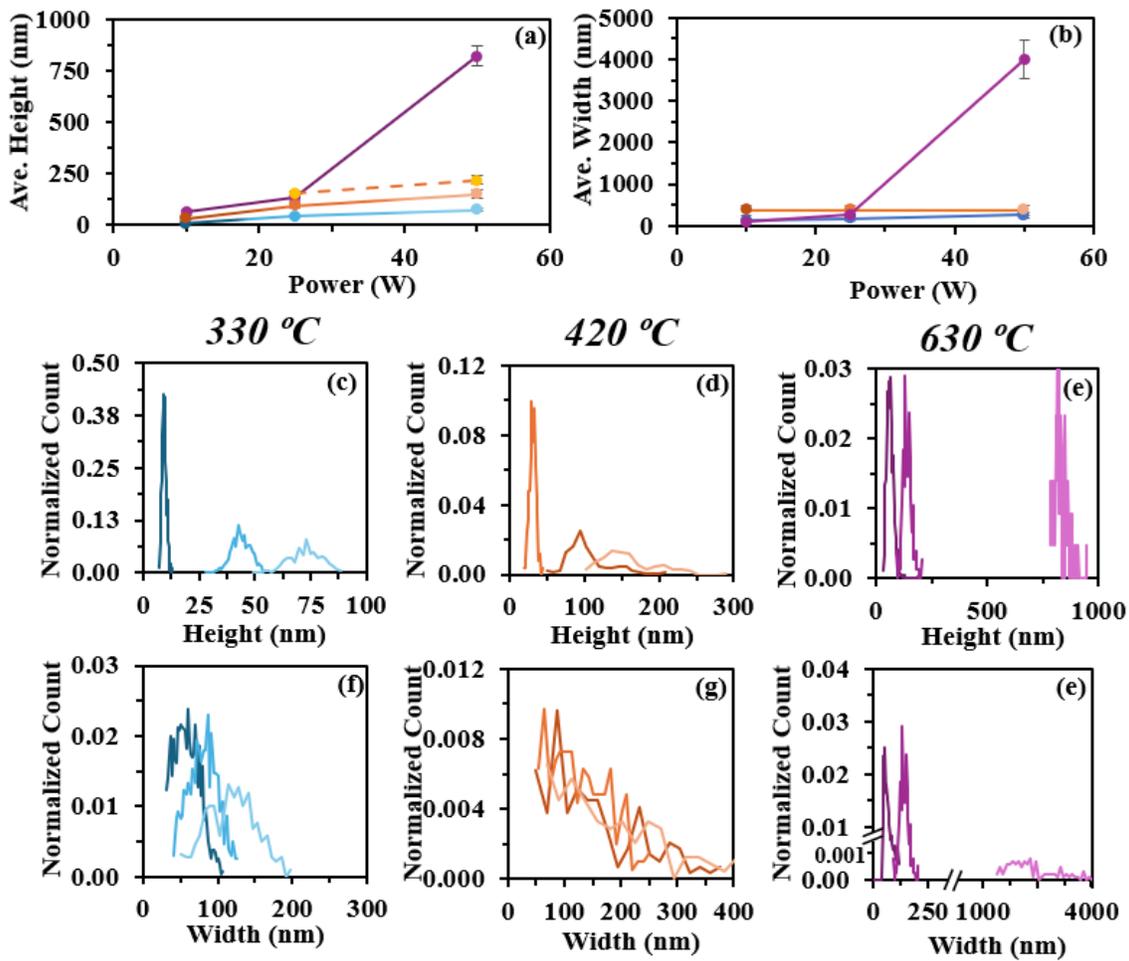

Figure 10 Grain analysis results from the AFM imaging in Figure 9 to ascertain the average grain size and height variation.

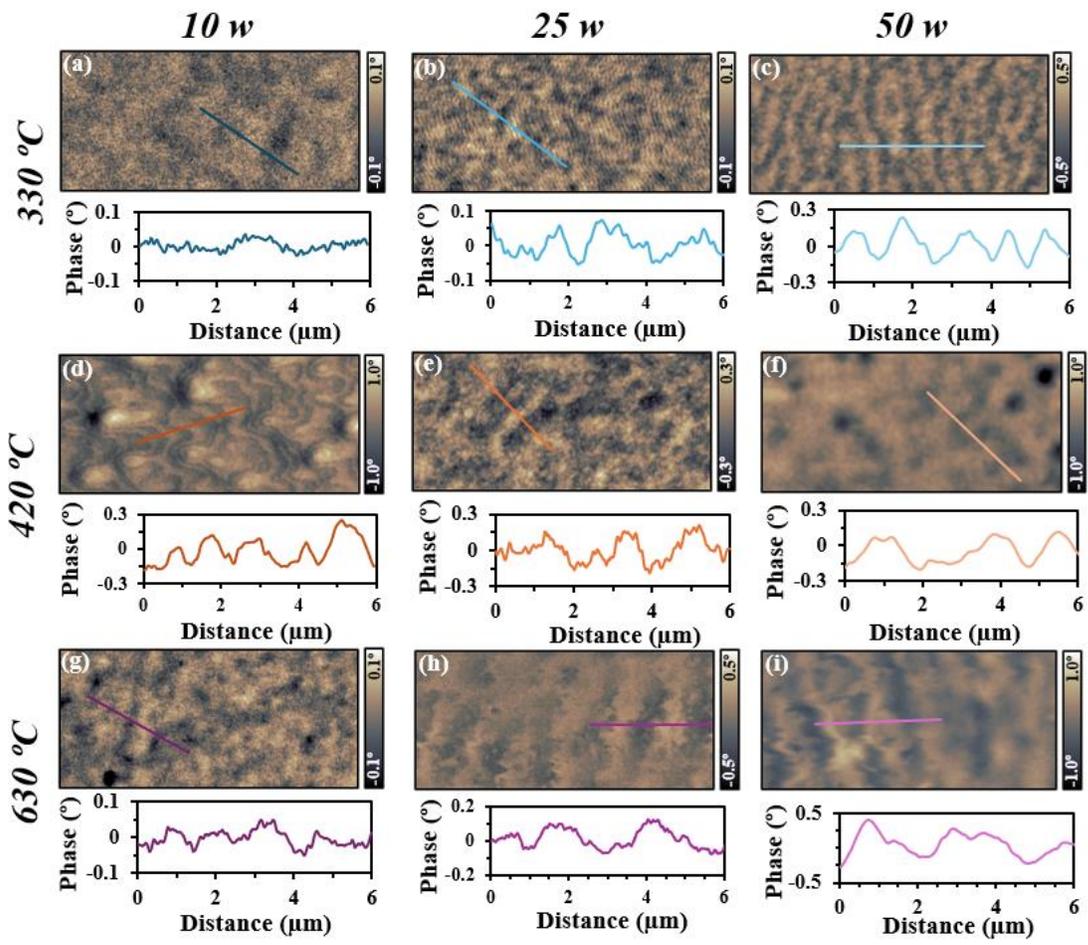

Figure 11 MFM Imaging of demagnetized samples showing the variation in domain size as a function of power and temperature. Line cuts show the variation in the domain size.

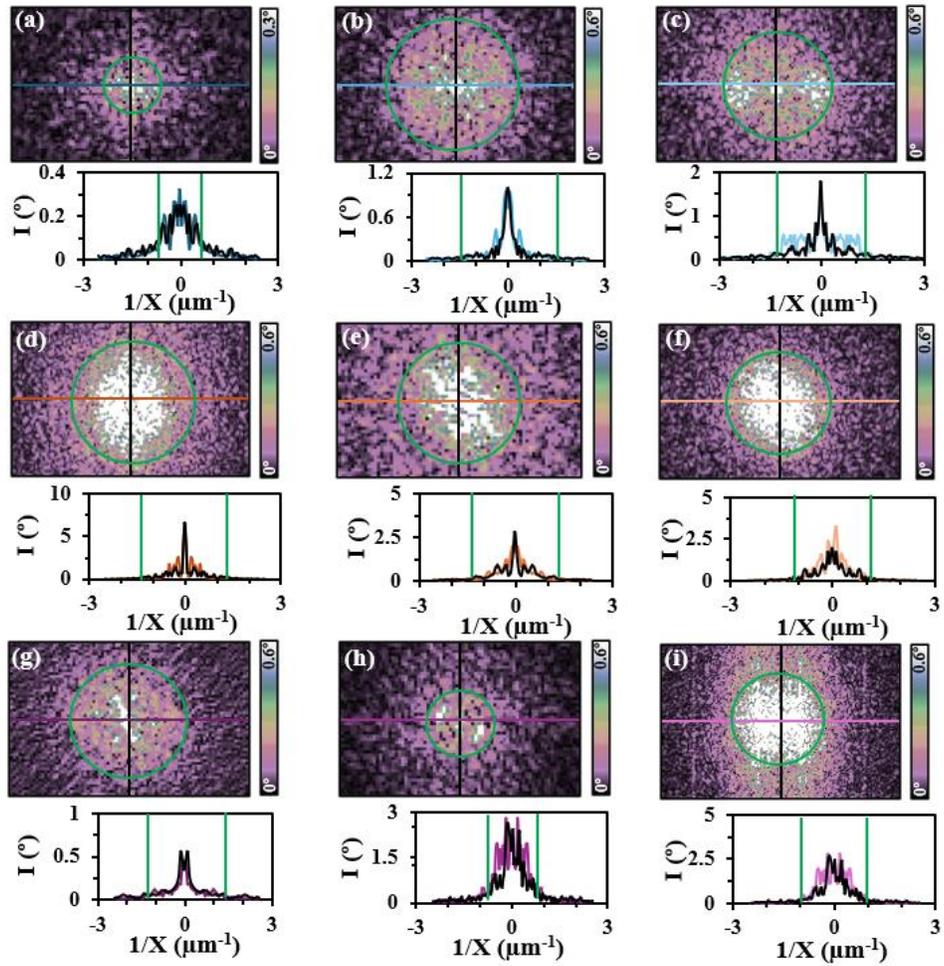

Figure 12 FFT of the domain structure Figure 11 to ascertain the average domain size.

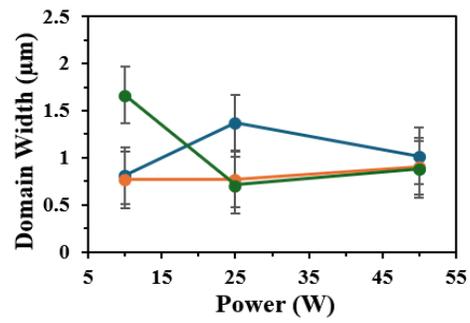

Figure 13 plots of domain size determined from FFT analysis as a function of sputtering power for the three temperature regimes.

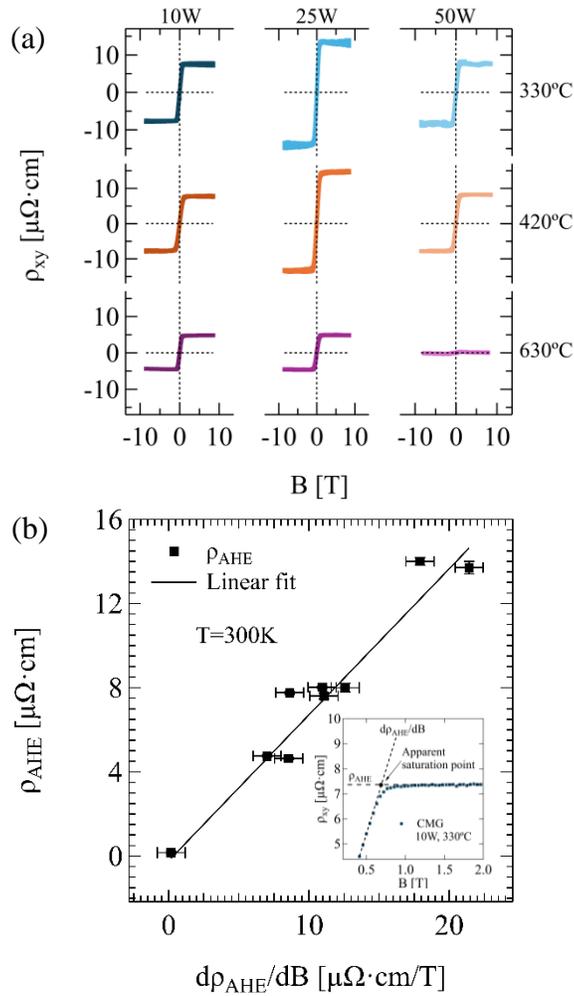

Figure 14 (a) Magnetic field B [T] dependence of transverse resistivity $\rho_{xy} = R_{xy} \cdot d$ [Ω·cm] of CMG grown at different RF powers and substrate temperatures. (b) The plot of the maximum Hall resistivity $\rho_{AHE}$ versus the zero-B slope $d\rho_{AHE}/dB$ (inset) obtained from each film. Dashed line : A linear fit to data ($0.694 \pm 0.066$ T slope, $-0.2 \pm 0.5$ μΩ·cm intercept)

# SUPPLEMENTARY MATERIAL

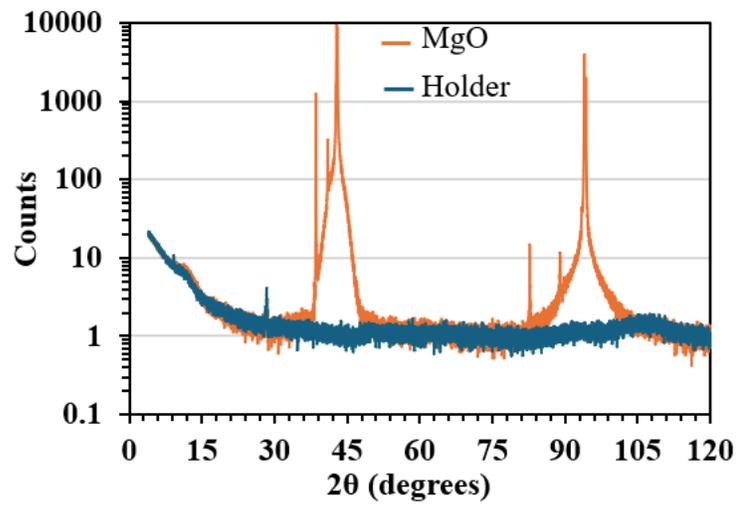

Figure S1 Symmetric scan of quartz zero-background sample holder (blue) and MgO (001) (orange) on the mounted on zero-background holder.